\documentclass{amsart}

\usepackage{amssymb}
\usepackage{amscd}

\begin{document}

\title{The Clausius inequality does not follow from
the second law of thermodynamics}

\author{Alexey~Gavrilov}
\email{gavrilov@lapasrv.sscc.ru}

\address{\newline Institute of Computational Mathematics and Mathematical
Geophysics;\newline ~Russia,~Novosibirsk}

\begin{abstract}
The example of macroscopic thermodynamical system violating
the Clausius inequality is presented.
\end{abstract}

\maketitle

\section{The gap in the proof of the Clausius inequality}

The well-known Clausius inequality
$$\oint\frac{dQ}{T}\le 0\eqno{(1)}$$
is supposed to be valid for any closed system undergoing  a cycle.
The more general formulation of this inequality
$$dS\ge\frac{dQ}{T}\eqno{(2)}$$
is actually less clear because the definition of entropy for nonequilibrum
systems is a subject of controversa. The rigirous version of (2) is
the so-called Clausius-Duhem inequality
$$\int_A^B\frac{dQ}{T}\le S_B-S_A,\eqno{(3)}$$
where $A$ and $B$ are {\it equilibrum} states.

The proof of (1) may be found in any textbook on thermodynamics.
Consider a system connected to a thermal energy reservour
at a constant absolute temperature of $T_B$
through a reversible cyclic device. It is also connected to a work
consumer (this system is necessary but usually not mentioned)
(fig. 1).

\begin{picture}(350,120)
\put(170,0){{\bf fig.1}}
\put(25,60){$T_B$}
\put(215,80){$T$}
\put(10,15){\framebox(40,100)}
\put(300,15){\framebox(40,100)}
\put(200,60){\framebox(40,40)}
\put(125,80){\circle{40}}
\put(50,80){\line(10,0){55}}
\put(145,80){\line(10,0){55}}
\put(70,84){$Q_B$}
\put(165,84){$Q$}
\put(270,84){$W_1$}
\put(270,44){$W_2$}
\thicklines
\put(240,80){\line(10,0){60}}
\put(125,60){\line(0,-10){20}}
\put(125,40){\line(10,0){175}}

\end{picture}

If the inequality (1) is violated then
$$\oint\frac{dQ}{T}=\frac{Q_B}{T_B}> 0,$$
where $Q_B$ is the amount of heat received by the reversible device
from the energy reservour. Due to the conservation of energy principle all this
heat is converted to the work. But the Kelvin-Plank statement of the
second law states that it is impossible to
take heat from one system, to give work to another system
and to make no changes in any other system. Thus we have a contradiction.

In this proof it is implicitely assumed that we may complete the cycle
without making any changes in any system except for the four systems
shown in fig.1. However, as we shall see, this assumption may be wrong.

\section{The adiabatic process should not be an isentropic one}

The {\it xenium} is a gas whose molecules may be in two different states
with the same energy levels. The spontaneous transitions between this states
are very rare. However two sufficiently close molecules may exchange
their states. It does'n matter whether the xenium exists. The question is
whether the existance of a gas with these properties violates some law of
nature or not. In author's opinion, it does not.

If in xenium the number of molecules in one of states
is more then in another state then we may consider this metastable gas as a
mixture of, say, xenium-1 and xenium-2. Let us consider the vessel
of xenium-1 and another one of equilibrum xenium
at the same temperature. Two vessels are separated by the membrane
which is thin and don't prevent molecules from interacting (fig. 2).

\begin{picture}(350,120)
\put(170,0){{\bf fig.2}}
\put(220,45){xenium-1}
\put(100,45){xenium}
\put(175,15){\line(0,10){70}}
\put(175,85){\line(1,1){10}}
\put(185,95){\line(10,0){70}}
\put(187,97){{\small thin membrane}}

\put(63,45){\circle{3}}
\put(90,74){\circle{3}}
\put(96,60){\circle{3}}
\put(102,22){\circle{3}}
\put(117,41){\circle{3}}
\put(128,67){\circle{3}}
\put(143,28){\circle{3}}
\put(154,76){\circle{3}}
\put(162,40){\circle{3}}

\put(55,21){\circle*{3}}
\put(66,65){\circle*{3}}
\put(80,37){\circle*{3}}
\put(113,67){\circle*{3}}
\put(121,27){\circle*{3}}
\put(124,62){\circle*{3}}
\put(152,54){\circle*{3}}

\put(188,35){\circle{3}}
\put(201,74){\circle{3}}
\put(223,58){\circle{3}}
\put(229,27){\circle{3}}
\put(239,41){\circle{3}}
\put(255,66){\circle{3}}
\put(268,24){\circle{3}}
\put(269,66){\circle{3}}
\put(287,40){\circle{3}}
\put(250,20){\circle{3}}
\put(180,43){\circle{3}}

\thicklines
\put(50,15){\line(0,10){70}}
\put(50,15){\line(10,0){250}}
\put(300,15){\line(0,10){70}}
\put(50,85){\line(10,0){250}}

\end{picture}

Then the equilibrum xenium will be enriched with xenim-1 and entropy
of this gas will {\it decrease} without any heat flow.
The only explanation of this paradox is the conclusion:
{\it entropy may be transferred from one system to another system adiabatically.}

Taking into account this process, the inequality (2) should be replaced by
$$dS\ge\frac{dQ}{T}+dI,\eqno{(4)}$$
where $dI$ is entropy absorbed by the system "by the direct way".
The inequality (1) takes the form
$$I+\oint\frac{dQ}{T}\le 0.\eqno{(5)}$$
If $I<0$ then (1) may be violated.

\section{Semiperpetuum mobile}

To build a system violating (1) we have to use some extra tricks.
The most simple is a selective membrane which is permeable for xenium-1
only (it may be replaced by a substance reacting with xenium-1, this is
less convenient but more realistic). This membrane is thick and don't
allow molecules to interact through it.

Our perpetuum mobile of the second kind is a cylinder with a piston
on one side, closed with a thin membrane on the other side.
The cylinder is divided
into two parts by a selective membrane. The volume between two membranes
is filled by xenium (fig. 3). This device has a straight connection to a
heat reservour and works isothermically. Besides, we have two large reservours
with xenium-1 one and with xenium-2 another.

\begin{picture}(350,150)
\put(170,0){{\bf fig.3}}
\put(300,15){\line(0,10){100}}
\put(225,60){xenium}
\put(120,60){xenium-1}

\put(300,115){\line(-1,1){10}}
\put(290,125){\line(-10,0){70}}
\put(225,127){{\small thin membrane}}

\put(175,115){\line(-1,1){10}}
\put(165,125){\line(-10,0){70}}
\put(100,127){{\small selective membrane}}

\thicklines
\put(175,15){\line(0,10){100}}
\put(100,18){\line(0,10){94}}
\put(90,18){\line(0,10){42}}
\put(90,112){\line(0,-10){42}}
\put(90,112){\line(10,0){10}}
\put(90,18){\line(10,0){10}}
\put(90,60){\line(-10,0){70}}
\put(90,70){\line(-10,0){70}}
\put(50,15){\line(10,0){250}}
\put(50,115){\line(10,0){250}}
\end{picture}

At the begining of the cycle the piston is pushed to the selective membrane
and the gas is in equilibrum.
Then the device is connected to the reservour with xenium-1 for some time,
so the gas in the cylinder becames enriched with xenium-1. Then the isothermic
expansion follows. After the expansion ends the device is connected
to the reservour with xenium-2 until the number of molecules in two
states {\it in all the cylinder} becames equal. The isothermic compression
completes the cycle.

Note that the working gas is the xenium-1 only. The expansion
takes place at higher (partial) pressure of xenium-1 then the compression.
Hence this device produces the work and consumes the heat, violating
the inequality (1). However, no contradiction to the second law arises
because entropy of reservours increases.

\end{document}